\documentclass[notitlepage,twocolumn,showpacs,preprintnumbers,amsmath,amssymb,floatfix,aps]{revtex4-1}

\usepackage{graphicx}% Include figure files
\usepackage{dcolumn}% Align table columns on decimal point
\usepackage{bm}% bold math
\usepackage{array} % for better arrays (eg matrices) in maths
\usepackage{color}
\usepackage{dsfont}
\usepackage{ulem}

\usepackage{pdfsync} 
\usepackage{multirow}
\usepackage{amsmath}
\usepackage{amsfonts}
\usepackage{mathrsfs}
\usepackage{enumitem}
\setlist{nolistsep} % or \setlist{noitemsep} to leave space around whole list
\usepackage[colorlinks=true, pdfstartview=FitV, linkcolor=blue, citecolor=blue, urlcolor=blue]{hyperref}

\newcommand  {\Rb} {$^{87}$Rb }

\newcommand{\ket}[1]{\left| #1 \right>} % for Dirac bras
\newcommand{\bra}[1]{\left< #1 \right|} % for Dirac kets
\begin{document}
%\preprint{APS/123-QED}

\title{Stueckelberg interferometry using periodically driven spin-orbit coupled Bose-Einstein condensates}% Force line breaks with \\

\author{Abraham J. Olson $^{1}$} 
\thanks{These authors contributed equally}
\author{David B. Blasing$^{1}$}
\thanks{These authors contributed equally}
\author{Chunlei Qu$^{2,3}$}
\author{Chuan-Hsun Li$^{4}$}
\author{Robert J. Niffenegger$^{1}$}
\author{Chuanwei Zhang$^{2}$}
\author{Yong P. Chen$^{1,4,5}$}
\thanks{yongchen@purdue.edu}

\affiliation{
$^{1}$ Department of Physics and Astronomy, Purdue University, West Lafayette, Indiana 47907, USA \\
$^{2}$ Department of Physics, The University of Texas at Dallas, Richardson, Texas 75080, USA \\
$^{3}$ INO-CNR BEC Center and Dipartimento di Fisica, Universit\`a di Trento, Povo 38123, Italy \\
$^{4}$ School of Electrical and Computer Engineering, Purdue University, West Lafayette, Indiana 47907, USA \\
$^{5}$ Purdue Quantum Center, Purdue University, West Lafayette, IN 47907, USA 
}

\date{\today}% It is always \today, today,
%             %  but any date may be explicitly specified
%

\begin{abstract}
We study the single-particle dispersion of a spin-orbit coupled (SOC) Bose-Einstein condensate (BEC) under the periodical modulation of the Raman coupling. This modulation introduces a further coupling of the SOC dressed eigenlevels, thus creating a second generation of modulation-dressed eigenlevels. Theoretical calculations show that these modulation-dressed eigenlevels feature a pair of avoided crossings and a richer spin-momentum locking, which we observe using BEC transport measurements. Furthermore, we use the pair of avoided crossings to engineer a tunable Stueckelberg interferometer that gives interference fringes in the spin polarization of BECs.
\end{abstract}
% PACS, the Physics and Astronomy
%                             % Classification Scheme.
%%\keywords{Suggested keywords}%Use showkeys class option if keyword
%                              %display desired

\maketitle
\section{Introduction}
In ultracold atoms, laser-induced synthetic gauge fields \cite{Lin_PRL_2009} have realized a rich variety of physics, such as synthetic electric \cite{Lin_NatPhys_2011} and magnetic \cite{Lin_Nature_2009} fields, spin-orbit coupling (also referred to as SOC) \cite{Lin_Nature_2011}, the superfluid Hall effect \cite{LeBlanc_PNAS_2012}, the spin Hall effect \cite{Beeler_Nature_2013}, and the Hofstadter/Harper and Haldane Hamiltonians \cite{Aidelsburger_PRL_2013,Miyake_PRL_2013,Jotzu_Nature_2014}. Many of these works use Raman-coupling between spin states of ultracold atoms to modify the single-particle dispersion relation \cite{Higbie_PRL_2002,Spielman_PRA_2009,Dalibard_RMP_2011}. This has resulted in a rich field of studies in one-dimensional (1D) equally weighted Rashba and Dressehauls SOC for both Bose-Einstein condensates and degenerate Fermi gases \cite{Zhang_PRL_2012, Cheuk_PRL_2012, Wang_PRL_2012, Zhang_PRA_2013stability, LeBlanc_NJP_2013, Hamner_Nature_2014, Fu_NatPhys_2014,Goldman2014,Zhai2015}. Such coupling has been combined with an optical lattice \cite{Hamner2015} and led to a softening of the roton and phonon modes \cite{Ji2015}. Furthermore, very recently, this synthetic SOC has been extended to 2D~\cite{Huang2016,Wu2016}, opening the door for the quantum simulation of various topological physics. 

In our previous work, we used BEC transport to study Landau-Zener (LZ) like transitions between the SOC dressed eigenlevels at the avoided crossings induced by a Raman coupling of constant strength \cite{Olson_PRA_2014}. Here we show that a modulation of the Raman coupling creates new SOC dressed band-structures, which we characterize by measurements of BEC transport and Landua-Zener transitions. Modulation of the Raman coupling was previously used to create a tunable SOC at high driving frequency \cite{Jimenez-Garcia2015}. When the driving frequency is instead comparable with the energy between the two dressed bands, the two bands couple together, inducing a richer spin-momentum locking and a pair of avoided crossings. In this work, we have experimentally observed both the richer spin-momentum locking and used the pair of avoided crossings to engineer a Stueckelberg \cite{Stueckelberg1932} interferometer. 

The remainder of this paper is structured as follows. In Sec. \ref{sec_expsetup}, we present our experimental setup and methods. In Sec. \ref{sec_expresults_groundband} we discuss our experimental results showing the difference in the spin momentum locking of the dressed ground band (created with a static Raman coupling) and the ``modulation-dressed" band (created with a periodically modulated Raman coupling). In Sec. \ref{sec_expresults_atom_interferometer} we show how we engineered a spin-resolved Stueckelberg atom interferometer using the pair of avoided crossings between the modulation-dressed bands. Finally, in Sec. \ref{sec_concl} we offer our concluding remarks and future prospects.

\section{Experimental setup and methods}
\label{sec_expsetup}

\begin{figure}[thb]
  \includegraphics[width=0.5\textwidth,trim= 0cm 0cm 12.5cm 0cm,clip=true]{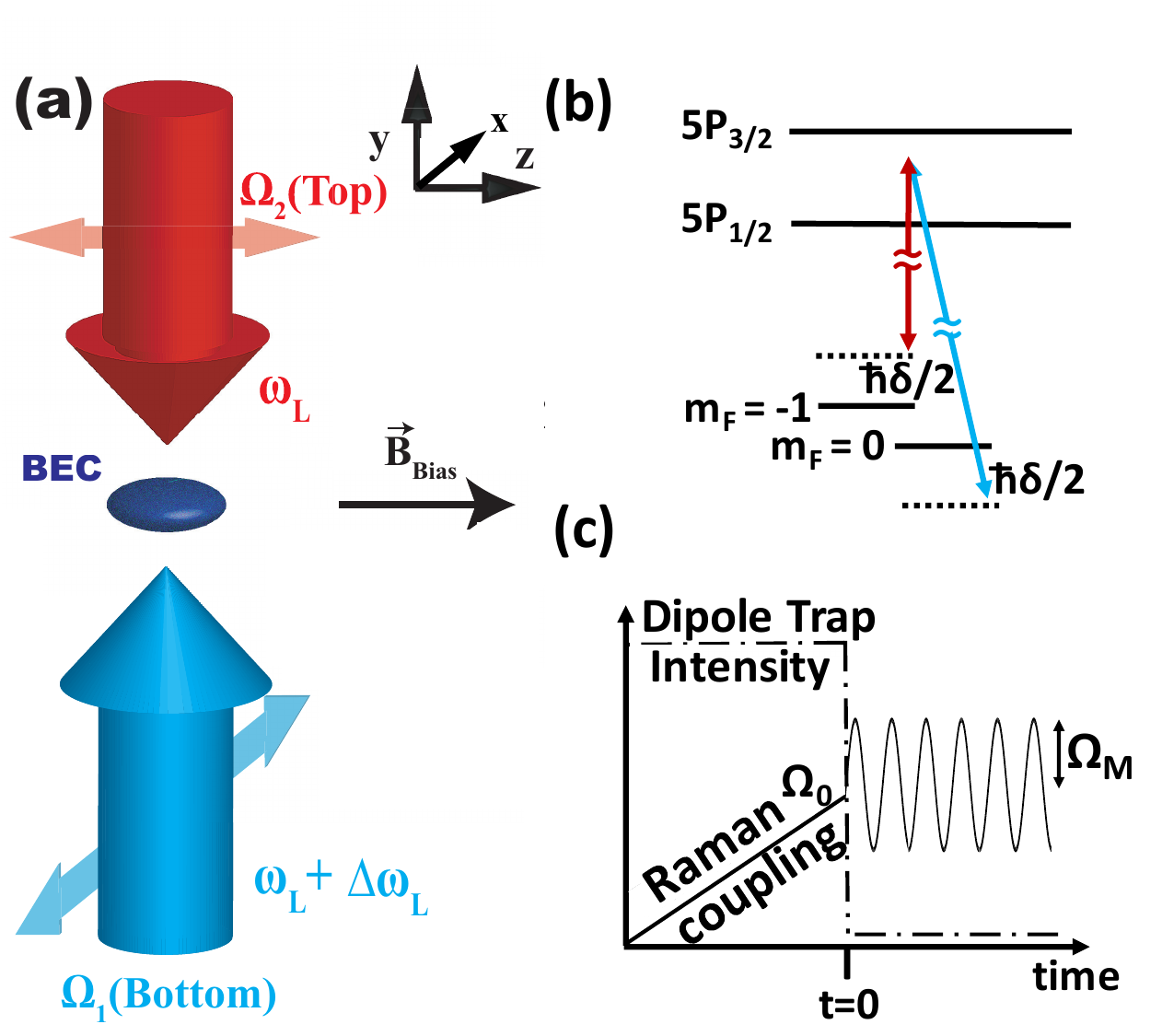}
  \caption{Experimental schematic. (a) Laser geometry showing both Raman beams. $\vec{B}_\text{Bias}$ denotes the bias magnetic field (of $\approx5G$) which lifts the degeneracy of the $m_F$ spin states. The acceleration due to gravity is along the $-\hat{y}$ direction. (b) Energy level diagram showing the two bare spin states, $\ket{F,m_F} = \ket{1,-1}$ and $\ket{1,0}$ (also denoted up and down respectively), and their Raman-induced coupling. Drawing is not to scale. (c) Representative timing diagram for the dipole trapping laser (dashed-dot line) and the Raman coupling (solid line).}
	\label{fig:expsetupdb}
\end{figure}

\begin{figure*}[t!]
  \includegraphics[width=.8\textwidth,trim= 0cm 5.5cm 11cm 12.5cm,clip=true]{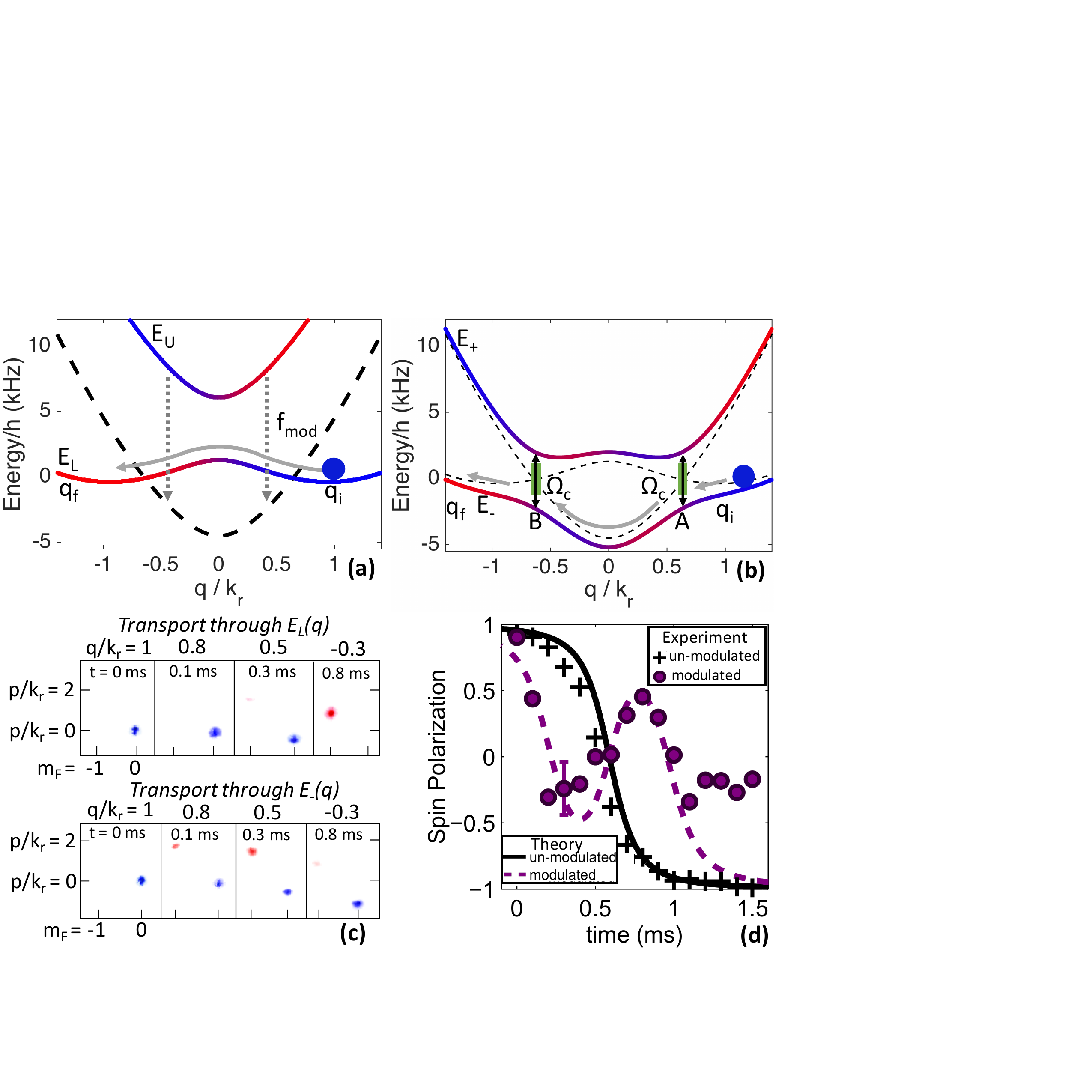}
  \caption{Experimental demonstration of the modified energy-momentum dispersion relation in the presence of modulated Raman coupling. (a) The unmodulated 1D SOC eigenlevels ($E_U(q)$ and $E_L(q)$) calculated from Eq. \ref{eqn:HSO} with $\Omega_0=1.3\, E_r$, and $\delta=0 \, E_r$. The dashed line shows the location of $E_U(q)$ if it had been shifted down by $f_{mod}=10.56$ kHz. (b) The $E_+(q)$ and $E_-(q)$ modulation-dressed eigenlevels calculated from Eq. \ref{eqn:H3pi} with identical $\Omega_0$ and $\delta$ as (a) and with $\Omega_C = 0.58 \,E_r$ and $f_{mod}=10.56$ kHz. The shifted but unmodulated dressed eigenlevels are shown by dashed lines. In (a) and  (b) the blue and red colors superimposed on the eigenlevels and images represent spin down (bare spin $m_F=0$) and up (bare spin $m_F=-1$) respectively. Note the richer spin-momentum locking of the modulation-dressed SOC eigenlevels and the two avoided crossings labeled A and B. The green bars indicate the avoided crossings with gapsize $\Omega_C$ used as beam splitters in the interference experiments discussed later. (c) and (d) Experimental comparison between the spin polarization of BECs transported through an unmodulated band $E_L(q)$ and a modulated band $E_-(q)$. Both bands used $\Omega_0=1.3\,E_r$ and $\delta=0\,E_r$. The modulated band had $\Omega_M=1.3 \,E_r$ and $f_{mod}=10.56$ kHz. The BECs started at $q_i\approx+ 1\,k_r$, fell under gravity with acceleration $\alpha_{F}\approx1680$ $k_r/s$ along the $-\hat{y}$ direction for $1.5$ ms, and reached $q_f\approx- 1.5\,k_r$. (c) Four representative time-of-flight images taken at quasimomentums $q/k_r$ of 1.0, 0.8, 0.5, and -0.3 for both $E_L(q)$ and $E_-(q)$ showing their different spin-momentum lockings. (d) Comparison between the observed and calculated spin polarizations for BECs in both $E_L(q)$ and $E_-(q)$. In the unmodulated case, the BEC nearly adiabatically follows the lowest energy eigenlevel and the expected monotonic spin rotation is observed (black crosses). However, in the modulated case, an additional oscillation of the spin polarization is observed (purple circles). Solid black (purple dashed) lines are the calculated spin polarization of $E_L(q)$ ($E_-(q)$) given the experimental parameters and $\Omega_C=0.58 \,E_r,$ using Eq. \ref{eqn:HSO}(\ref{eqn:H3pi}). For this and following figures, a representative error bar indicates an average of 10\% uncertainty in atom population in each spin due to technical noise.}
	\label{fig:expSetup}
\end{figure*}

Our experimental setup used to create a 1D Raman-induced SOC is shown in Fig. \ref{fig:expsetupdb}. Many details of our apparatus are contained in Refs.~\cite{Olson_PRA_2013,Olson_PRA_2014}. The Raman beams (whose beam waist is large compared to the in-situ size of the BEC) counterpropagate along the same axis as gravity ($\hat{y}$). Thus by reducing the intensity of the dipole trapping laser, the BEC can be accelerated by gravity along the $-\hat{y}$ direction. In all the following experiments, we use this technique to induce BEC transport through both the dressed and modulation-dressed bandstructures induced by the Raman beams.

Our experiment starts with a 1D SOC BEC of \Rb atoms subjected to a constant Raman coupling of two spin states ($|\uparrow\rangle =|F=1,m_F=-1\rangle$ and $|\downarrow\rangle=|F=1,m_F=0\rangle$), as shown in Fig.~\ref{fig:expsetupdb} (a) and (b). This coupling creates two eigenlevels, both of which possess a spin-momentum locking in the quasimomentum ($\hbar q$) space (the total spin polarization of the eigenstate depends on $q$). We denote the upper and lower eigenlevels as $E_U (q)$ and $E_L(q)$. This unmodulated eigenlevel structure, an example of which is pictured in Fig. \ref{fig:expSetup} (a), is calculated from the SOC Hamiltonian

\begin{equation}\label{eqn:HSO}
{\cal H}_{SOC}/\hbar = \left( \begin{array}{cc}
\frac{\hbar}{2 m} (q+k_r)^2 - \delta/2 & \Omega_0/2 \\
\Omega_0/2 & \frac{\hbar}{2 m} (q-k_r)^2 + \delta/2 \end{array} \right),
\end{equation}

\noindent where $\delta$ is the Raman laser detuning from the energy difference between spin states due to the Zeeman effect ($\delta$ is zero in all these experiments), $m$ is the \Rb atomic mass, $\hbar k_r = h/\lambda_R$ is the recoil momentum of the Raman laser with wavelength $\lambda_R=790$ nm, $\hbar=h/2\pi$ is the reduced Planck's constant, and $\Omega_0$ is the unmodulated Raman coupling. For the remaining experiments, we define the total spin polarization of the BEC as $\mathcal{S}=(N_{\downarrow}-N_{\uparrow})/(N_{\downarrow}+N_{\uparrow})$, where $N_{\uparrow (\downarrow)}$ is the number of spin up (down) atoms in the BEC.  The recoil energy from the Raman lasers is $E_r=\hbar^2 k_r^2/2m= h \times 3.68$ kHz. ${\cal H}_{SOC}$ only includes the $m_F=-1$ and $0$ states since the $m_F=+1$ spin state is far detuned in the range of $q$ accessed in these experiments because of the quadratic Zeeman shift and the recoil energy associated with the two-photon Raman transfer.

To engineer a new dispersion relation for our ultracold atoms, we added a time-dependent modulation to the intensity of the Raman-coupling: $\Omega_R(t)=\Omega_0+\Omega_M \cos(2 \pi f_{mod} t)$, where $f_{mod}$ is the modulation frequency and $\Omega_M$ is the modulation amplitude, see the timing diagram in Fig. \ref{fig:expsetupdb} (c). Two results demonstrated the creation of the new modulation-dressed eigenlevels: (i) we observed the more complex rotation of the spin polarization of BECs during transport through the modulation-dressed band (results shown in Fig. \ref{fig:expSetup}), and (ii) we used the pair of avoided crossings available in the modulation-dressed band to engineer an atom interferometer (results shown in Figs. \ref{fig:diffOmegaR}, \ref{fig:varydqdt}, and \ref{fig:varyOmegaM}). 
 
According to Floquet theorem (see Ref. \cite{Eckardt2015} for a recent discussion), periodically driven quantum systems can be described by Floquet states and a quasi-energy spectrum. The latter can be obtained by diagonalizing the following block tridiagonal matrix:
\begin{equation}
{\cal H}=
\left(
\begin{array}{ccccc}
\ddots & &  &  &  \\
 & \mathcal{H}_{SOC}+hf_{mod}\mathds{1}_2 & \mathcal{V}_{+1} & &  \\
 & \mathcal{V}_{-1} & \mathcal{H}_{SOC} & \mathcal{V}_{+1} &  \\
 &  & \mathcal{V}_{-1} & \mathcal{H}_{SOC}-hf_{mod}\mathds{1}_2 &  \\
 &  &  & & \ddots
\end{array}
\right)
\label{eq:Hblock}
\end{equation}

\noindent where $\mathcal{H}_{SOC}$ is the unmodulated SOC Hamitonian defined in Eq. \ref{eqn:HSO} and ${\cal V}_{\pm 1}=T^{-1}\int_0^T \mathcal{H}_{SOC}(t)e^{\pm i2\pi f_{mod}t} dt=
(\Omega_M/4)\sigma_x$ with $T=1/f_{mod}$ denoting the period of the external driving, $\sigma_x$ is the Pauli matrix, $\mathds{1}_2$ is the $2\times2$ identity matrix and $\mathcal{H}_{SOC}(t)$ is the same as $\mathcal{H}_{SOC}$ but replacing $\Omega_0$ with $\Omega_R(t)$. The eigenenergy spectrum of $\mathcal{H}$ exhibits a periodic pattern of the form $E_{\pm}(q)+n2\pi f_{mod}$ where $n=\pm 1, \pm 2, \ldots$. We call $E_{\pm}(q)$ the upper and lower modulation-dressed bands. The time-avaraged dynamics of the driven system can be well described by such modulation-dressed bands. Furthermore, when $hf_{mod}$ is slightly larger than $E_{U}(q \approx 0)-E_{L}(q \approx 0)$, $E_+(q)$ and $E_-(q)$ also feature a double avoided crossing with a gap size $\Omega_C$ (see Fig. \ref{fig:expSetup} (b)). The relationship between $\Omega_C$ and $\Omega_M$ can be calculated numerically and will be discussed later in Fig.~\ref{fig:varyOmegaM}(c).

However, we also found that a simpler, perhaps more intuitive, two-by-two Hamiltonian ${\cal H}_{mod}$ sufficiently explains our data in the parameter regimes studied \footnote{This further approximation becomes less accurate for smaller $f_{mod}$ and/or larger values of the quasimomentum than what we used.}. The two modulation-dressed bands $E_\pm(q)$ can be approximately modeled by coupling the lower dressed band $E_{L}(q)$ and the downshifted higher dressed band, $E_{U}(q)-hf_{mod}$, with a simple effective coupling constant $\Omega_C$, i.e., 

\begin{equation}\label{eqn:H3pi}
{\cal H}_{mod} = \left( \begin{array}{cc}
E_{L} (q) & \hbar\Omega_C/2 \\
\hbar\Omega_C/2 & E_{U}(q)-hf_{mod} \end{array} \right).
\end{equation}
Diagonalizing $\mathcal{H}_{mod}$ at each $q$, we obtain the new modulation-dressed eigenlevels (an example of which is shown in Fig. \ref{fig:expSetup} (b)). These modulation-dressed eigenlevels of Eq. \ref{eqn:H3pi} are nearly identical with $E_{\pm}(q)$ in the parameter regime of our experiments and we used them in our following analysis \footnote{In the parameter regimes that we used, we calculate the difference in the spin polarization between the eigenlevels predicted by $\mathcal{H}_{mod}$ and $\mathcal{H}$ to be less than $\approx4\%$, which is less than our experimental error.}. The modulation-dressed bands $E_+(q)$ and $E_-(q)$ feature a more complex rotation of the spin polarization of the BEC as the quasimomentum goes from $+\hbar k_r$ to $-\hbar k_r$. This contrasts with the monotonic single rotation of the spin polarization present in $E_U(q)$ and $E_L(q)$, see Fig. \ref{fig:expSetup} (a) and (b).

\section{Spin momentum locking of the unmodulated and modulated bands}
\label{sec_expresults_groundband}

To study the spin composition of the modulation-dressed band $E_-(q)$, we use the BEC transport method developed in our earlier work \cite{Olson_PRA_2014}. Briefly, a BEC is initially prepared in a bare $m_F=0$ state, and subsequently it is adiabatically loaded at $q_i \approx 1 k_r$ of a dressed band with a fixed value of $\Omega_0$. The modulation of the Raman beams is then turned on at the same time (defined as $t=0$) as the optical trap holding the BEC is lowered or turned off, which allows gravity to accelerate the BEC in the $-\hat{y}$ direction at a tunable average rate $\alpha_{F}$ through both the avoided crossings of $E_-(q)$ (labeled A and B in Fig.~\ref{fig:expSetup} (b)). The probability of a diabatic transition between the modulation-dressed eigenlevels is given by the Landau-Zener formula, $P_{LZ} = \exp \left[-2 \pi(\Omega_C/2)^2/(\hbar \alpha \beta)\right]$, where $\alpha = |dq /dt|$ is the rate of acceleration at the avoided crossing and $\beta$ is the difference of the slopes of the unmodulated SOC energy levels. Although $P_{LZ}$ is an approximate formula to describe the probability of the transition between the energy bands in our experiment, it provides a reasonable and intuitive explanation of both our previous \cite{Olson_PRA_2014} and current work. After passing through both avoided crossings, the Raman beams and any remaining portion of the dipole trap are turned off instantaneously and the BEC is imaged after 15 ms of time-of-flight expansion, during the later portion of which, a Stern-Gerlach field is applied to separate the $m_F$ spin components.

Figure~\ref{fig:expSetup} (c) and (d) show time-of-flight images and spin polarizations respectively of BECs traversing along $E_L(q)$ and $E_-(q)$, revealing the different spin-momentum locking in these two ground dressed bands (without and with modulation). For $E_L(q)$ we used $\Omega_0=1.3\,E_r$, $\delta=0\,E_r,$ and $\Omega_M=0\,E_r$. For $E_-(q)$ we used the same $\Omega_0$ and $\delta$ but $\Omega_M$ was $1.3\,E_r$ and $f_{mod}$ was $10.56$ kHz. All BECs were accelerated by gravity ($\alpha_F =1680$ $k_r/$s = 9.8 m/s$^2$ along $-\hat{y}$). In panel (c), we show representative time-of-flight images at quasimomenta ($1.0,0.8,0.5,$ and$ -0.3\,k_r$) that highlight the difference in the spin composition and the spin momentum locking between $E_L(q)$ and $E_-(q)$, shown in the upper and lower rows respectively. Panel (d) shows the extracted spin polarization along both $E_L(q)$ and $E_-(q)$. In $E_L(q)$, the measured BEC spin polarization (black crosses) follows the calculated spin polarization (black line). When a strong modulation of $\Omega_M=1.3\,E_r$ is applied, the BEC instead exhibits (purple circles) the distinct spin polarization of $E_- (q)$ (purple dashed line). This modulation was strong enough to open a sufficient gap $\Omega_c$ so that the probability for non-adiabatic inter-eigenlevel transitions in the modulated-induced band structure were reasonably small ($P_{LZ}=0.14$). Such small non-adiabatic inter-eigenlevel transitions ensured that the spin polarization of the BECs were dominated by the lower band $E_- (q)$. However, the measured spin polarization of the BEC does not perfectly match the calculated spin polarization of $E_-(q)$ after about $1$ ms. We attribute this to the imperfect loading into the modulation-dressed bands and the weak but not completely negligible non-adiabatic inter-eigenlevel transitions \cite{Olson_PRA_2014}. Nonetheless, this experiment demonstrates the viability of modulated-Raman coupling to create a more complicated spin-momentum locking in $E_-(q)$, which is different from the previously studied ground band $E_L(q)$ induced by a static Raman coupling, and may offer new possibilities to explore spinor BEC physics. 

\section{Engineering a spin-resolved Stueckelberg atom interferometer}
\label{sec_expresults_atom_interferometer}

\begin{figure}[thb]
  \includegraphics[width=0.5\textwidth,trim= 0cm 2cm 9cm 0cm,clip=true]{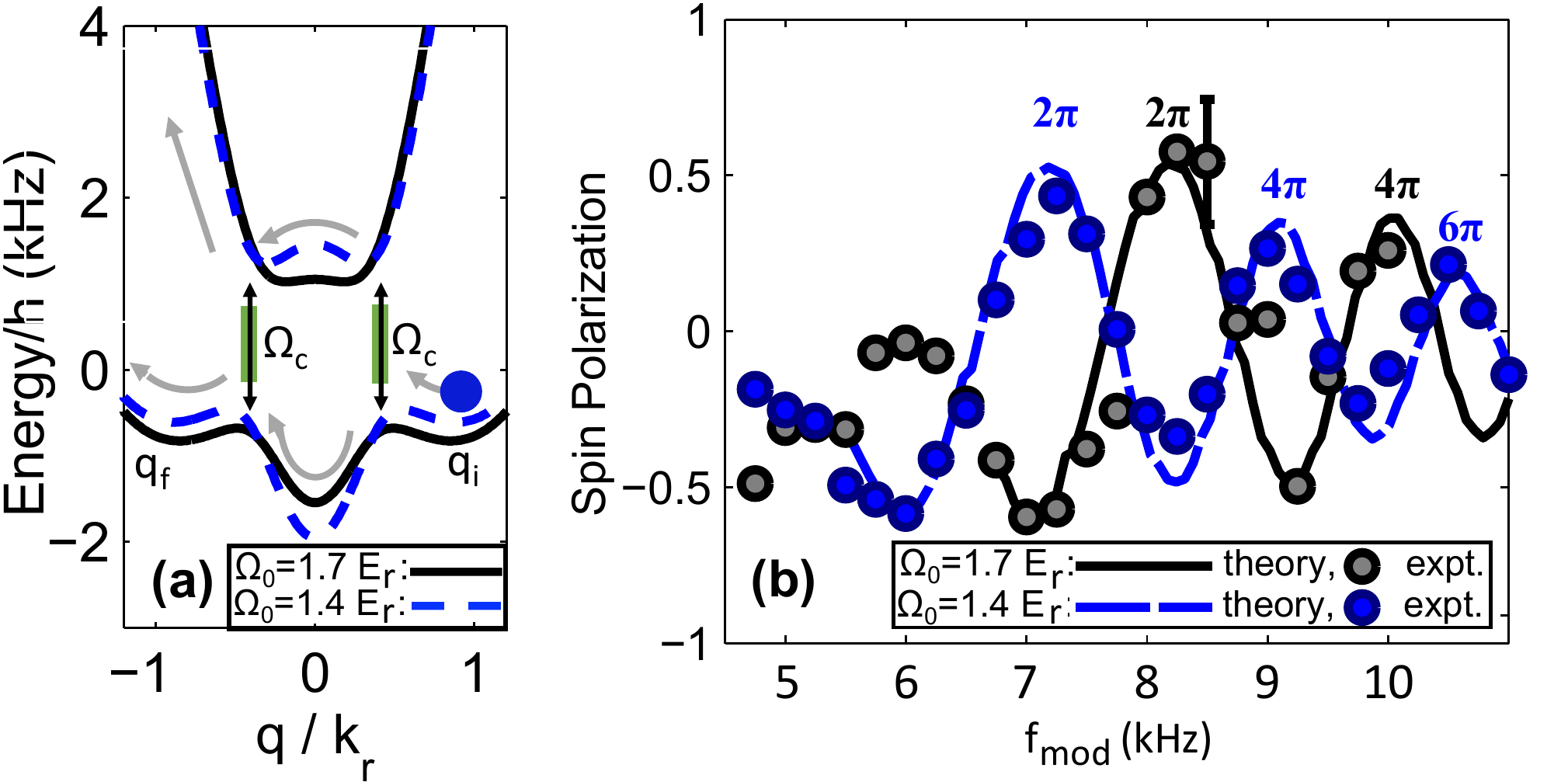}
  \caption{Measurement of Stueckelberg interference (a) Two representative modulation induced spin-orbit eigenlevels with $\Omega_0=1.4\,E_r$ (blue dashed line) and $1.7\,E_r$ (black solid line). Both eigenlevel calculations used $\Omega_C=0.3\,E_r$, $f_{mod}=8$ kHz, and $\delta=0\, E_r$. (b) Measured Stueckelberg interference fringes in the BEC spin polarization vs $f_{mod}$ at $\alpha_F=1680\,k_r/s$ for $\Omega_M=0.7 \,E_r$ (blue) and $0.8\,E_r$ (black). The theoretical curves were calculated from Eq. \ref{eq:Stot} with $\Omega_C=0.3\,E_r$, $\sigma_{\alpha}=0.07 \alpha_F$, $f_{np}=0.4$, $\delta=0$ and the same $\Omega_0$ as in the experiment (1.7$\,E_r$ for black and 1.4$\,E_r$ for blue).}
	\label{fig:diffOmegaR}
\end{figure}

In addition to studying the more complex spin-momentum locking of the modulation-dressed band, we also used the pair of avoided crossings between $E_+(q)$ and $E_-(q)$ to engineer an atom-interferometer. (Such a pair of avoided crossings is not realized in dressed eigenlevels created by unmodulated Raman coupling, see Fig.~\ref{fig:expSetup}(a) \footnote{We note in the three-state picture there are a pair of avoided crossings between the lowest two eigenlevels. However,  diabatic transitions to the highest eigenlevel are favored in this pair due to the small gap between the second and third eigenlevels and thus cannot function as an interferometer.}.) Stueckelberg interference \cite{Cronin_RMP_2009,Kling_PRL_2010} can occur upon the recombination of a wavefunction that was split along different energy eigenbands. By traveling along different eigenbands, each component may acquire a different phase. We observed such interference in our experiment after the following sequence of events (as depicted in Figs. \ref{fig:expSetup} (b) and \ref{fig:diffOmegaR} (a)). First, the BEC was coherently split into two components via a LZ transition at the avoided crossing A (labeled in fig. \ref{fig:expSetup} (b)); one component along $E_+(q)$ and the other along $E_-(q)$. Second, the components separately traveled along $E_+(q)$ and $E_-(q)$ and thus acquired a different phase. Finally, the two components recombined and interfered after another LZ transition at the avoided crossing B. The final spin composition of the BEC depends on the difference of the phase accumulated by each component while traversing $E_+(q)$ and $E_-(q)$; this phase difference depends on the energy difference between those paths and the time it takes to traverse them. If either the path or transport time is varied, the final spin polarization of the recombined BEC will change. The difference in energy between these bands as well as the separation in $q$ space between the beam splitters (i.e. the avoided crossings) are tunable via $f_{mod}$, and the transport time is controlled by $\alpha_F$. We define $\Phi$ (which is sometimes referred to as the Stueckelberg phase) as the total phase difference acquired between the two components of a BEC traveling separately along $E_+ (q)$ and $E_-(q)$. This phase difference for an atom with transport induced by a specific acceleration $\alpha$ is

\begin{equation}
\label{eqn:PhaseDiff}
 \Phi(\alpha)=\int_{q_{A}}^{q_{B}} \left[ E_+(q)-E_-(q) \right] dq / (\hbar \alpha).
 \end{equation}
 
The output spin polarization, $\mathcal{S}$, for an atom moving through this interferometer with acceleration $\alpha$ is calculated as (see appendix):

\begin{equation}
\label{eqn:SpinPolTheory}
\mathcal{S} (\alpha)=4\left[P_{LZ}(\alpha)-P_{LZ}(\alpha)^2\right] \cos \left[\Phi(\alpha)\right]-[1-2 P_{LZ}(\alpha)]^2
\end{equation}

\begin{figure}[hbt]
  \includegraphics[width=0.48\textwidth,trim= 0cm 0cm 9.1cm 0cm,clip=true]{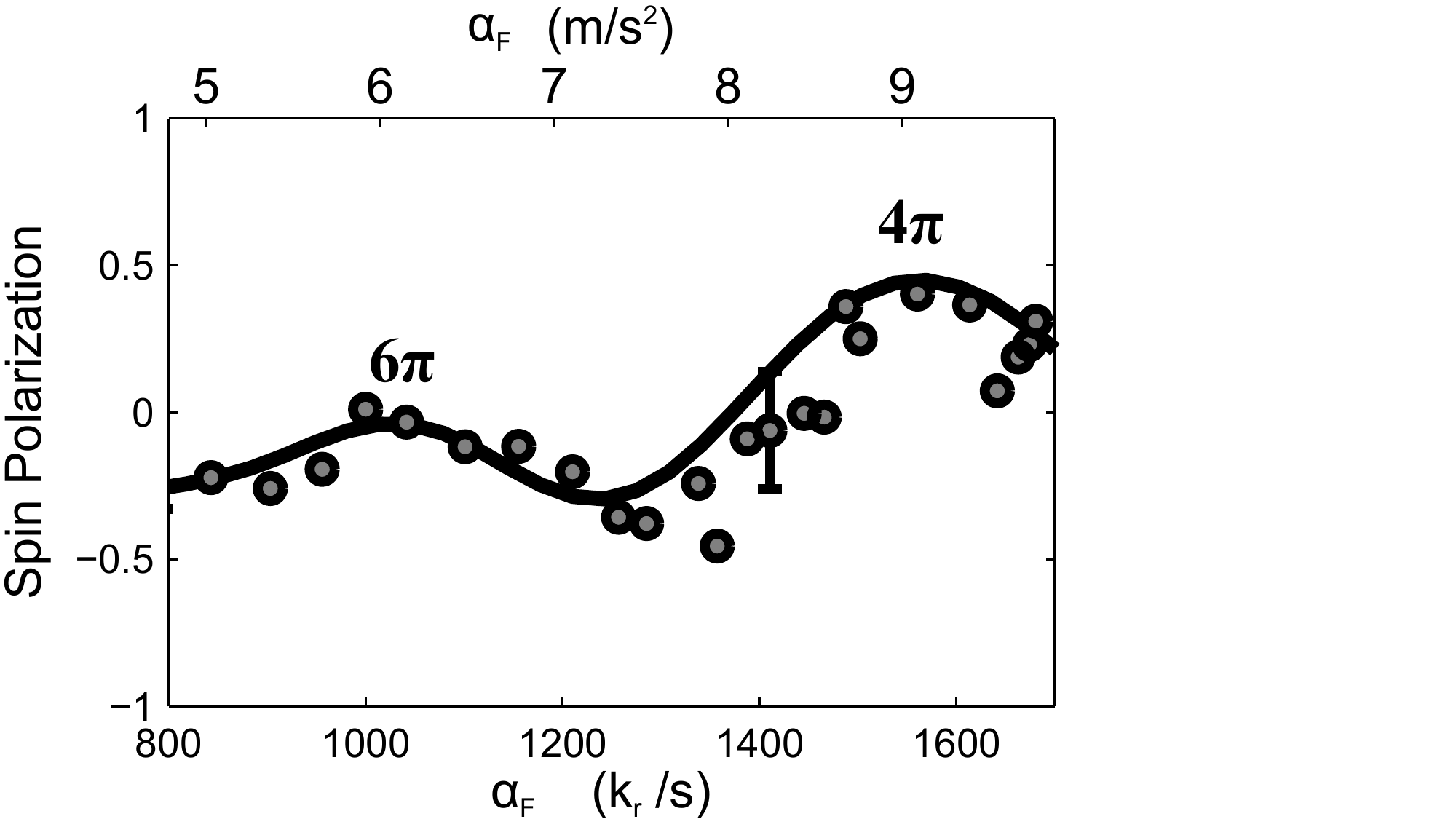}
  \caption{Stueckelberg interference fringes in the spin polarization of BECs at various values of the initial BEC acceleration, $\alpha_F$. Varying $\alpha_F$ changes the transport time through $E_+(q)$ and $E_-(q)$. This experiment used $\Omega_M=0.7\,E_r$, $\Omega_0=1.4\,E_r$, $f_{mod}=8.5$ kHz, and $\delta=0\, E_r$. The theoretical curves were calculated from Eq. \ref{eq:Stot} and used $\Omega_C=0.33\,E_r$, $\sigma_{\alpha}=0.07 \times (1680$ $k_r/s)$, and $f_{np}=0.3$. (The fringe contrast is strongly reduced at smaller $\alpha_F$ as the dephasing effect of $\sigma_{\alpha}$ gets larger with longer total time.)}
	\label{fig:varydqdt}
\end{figure}

\begin{figure*}[htb]
  \includegraphics[width=\textwidth,trim= .2cm 4.2cm 1.7cm 3.6cm,clip=true]{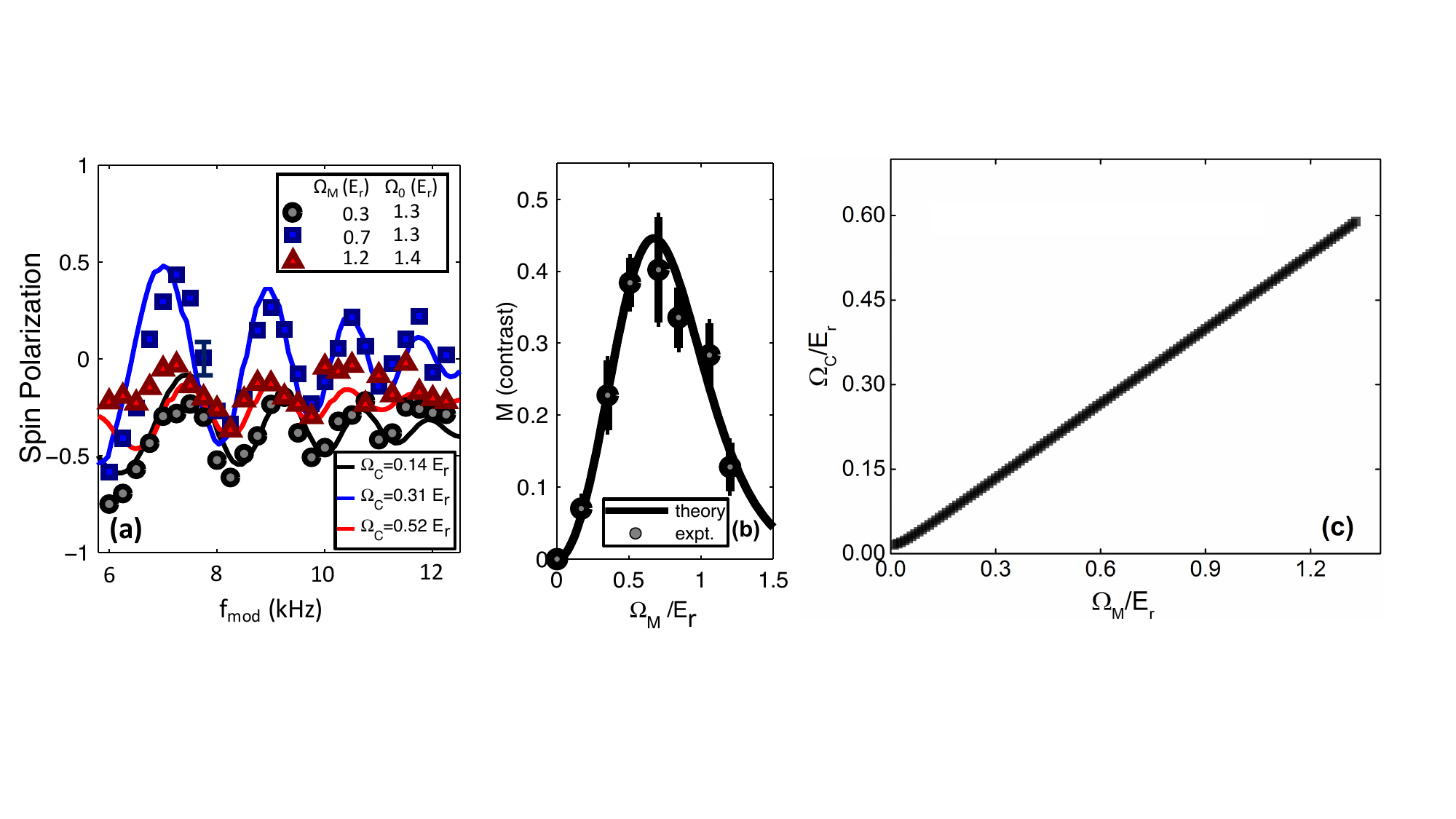}
  \caption{Effect of $\Omega_M$ on Stueckelberg interference. (a) Observed Stueckelberg interference fringes for $\Omega_M=0.3\,E_r$ and $\Omega_0 = 1.3\,E_r$, and $\Omega_M=0.7\,E_r$ and $\Omega_0 = 1.4\,E_r$, and $\Omega_M=1.2\,E_r$ and $\Omega_0 = 1.3\,E_r$ for black circles, blue squares, and red triangles; all used $\delta=0\,E_r$ and $\alpha_F=1680 \,k_r/s$. All the theoretical curves were obtained from Eq. \ref{eq:Stot} with $\delta=0\,E_r$, $\sigma_{\alpha}=0.07 \alpha_F$ and the experimental values of $\Omega_0$.  The black, blue, and red lines used $\Omega_C={0.14,0.31,0.52}\, E_r$ and $f_{np}={0.4,0.45,0.5}$ respectively. (b) Measured fringe contrast $M$ versus modulation amplitude $\Omega_M$. The theory curve was generated assuming a linear relation of $\Omega_C=\Omega_M/2.25$, $\Omega_0=1.4 \,E_r$, $f_{np}=0.4$, and $\sigma_{a}=0.07 \alpha_F$. Error bars indicate numerical fitting uncertainty of the fringes. (c) Theoretical calculation of $\Omega_C$ vs $\Omega_M$ with $\delta=0\,E_r$ and $\Omega_0=1.33\,E_r$. A linear fit gives $\Omega_C=\Omega_M/2.27$. This calculated ratio between $\Omega_C$ and $\Omega_M$ is found to change by less than $10\%$ in the range $\Omega_0$ experimentally accessed.}
	\label{fig:varyOmegaM}
\end{figure*}

We verified the operation of our modulation induced Stueckelberg interformeter in $\mathcal{S}$ in three separate experiments. (The interference fringes in $\mathcal{S}$ are due to the $\cos \left(\Phi(\alpha)\right)$ term in eq. \ref{eqn:SpinPolTheory}.) In the first two experiments, we saw Stueckelberg interference fringes while tuning $\Phi$ by separately changing $f_{mod}$ and $\alpha_F$. Then, in the final experiment, we observed the contrast of the spin polarization for different values of $\Omega_M$. The results from these three experiments are respectively shown in Figs. \ref{fig:diffOmegaR}, \ref{fig:varydqdt}, and \ref{fig:varyOmegaM}. 

First, at various values of $f_{mod}$, we measured Stueckelberg interference fringes using eigenlevel structures similar to those shown in Fig.~\ref{fig:diffOmegaR}(a). When the driving frequency $f_{mod}$ is varied, so are the energy difference $E_+(q)-E_-(q)$ and the ``lengths" of the interferometer arms in $q$-space. Consequently, according to Eq.~\ref{eqn:PhaseDiff}, the phase difference $\Phi$ is changed, which alters the spin polarization $\mathcal{S}$ according to Eq.~\ref{eqn:SpinPolTheory}. Figure ~\ref{fig:diffOmegaR} (b) shows the measured spin polarization of the BEC after it has passed through both avoided crossings with labels for the calculated phase differences of $\Phi=2\pi$, $4\pi$ and $6\pi$. This experiment was run at both $\Omega_0=1.4$ and $1.7\,E_r$, and the diagram Fig.~\ref{fig:diffOmegaR}(a) shows that the smaller $\Omega_0$ had a greater energy separation between its two modulation-dressed eigenlevels. This was reflected in the interference fringes in Fig.~\ref{fig:diffOmegaR}(b): to reach the same $\Phi$,  $\Omega_0=1.4E\,_r$ required a smaller $f_{mod}$ (i.e. a smaller $q_A-q_B$) as compared to  $\Omega_0=1.7\,E_r$.

The second method we used to tune $\Phi$ was instead at a fixed $f_{mod}$, but different times during which the BEC traveled along $E_+(q)$ and $E_-(q)$. This transport time was varied by changing the initial average acceleration of the BEC. The resulting Stueckelberg interference fringes as a function of $\alpha_F$ is shown in Fig.~\ref{fig:varydqdt}. An optical dipole trapping force was applied to reduce $\alpha_F$ relative to that caused by gravity, and thus increase the time that the BEC took to traverse the two energy paths (a similar technique was used in Ref. \cite{Olson_PRA_2013}). Accelerations that caused calculated phase accumulations $\Phi=4\pi$ and $6\pi$ are labeled. The reduced contrast at smaller $\alpha_F$ is due to the increased time for interactions to broaden the velocity distribution of the BEC, and thus dephase the BEC as it traverses $E_+(q)$ and $E_-(q)$. However, the fringes are still apparent and the model agrees well with the experimental results.

A fraction of the atoms did not participate in the Stueckelberg interference in our experiments, likely due to non-adiabatic initial state preparation in the modulation-dressed band. This fraction is treated as a fitting parameter $f_{np}$. In addition, the BEC may experience a non-uniform acceleration distribution about $\alpha_{F}$ due to atom-atom interactions. The non-uniform acceleration is assumed to follow a Gaussian distribution, $n(\alpha)=\frac{1}{\sqrt{2\pi}\sigma_{\alpha} } \exp \left[-(\alpha-\alpha_F)^2/(2 \sigma_{\alpha}^2) \right]$. The values used in this paper for $\sigma_{\alpha}$ are consistent with numerically calculated solutions of the Gross-Pitaevski equation using a variational method with Gaussian ansatz and parameters similar to these experiments \cite{Olson_PRA_2013a}. Accounting for $n(\alpha)$ and the non-participating fraction, the total spin polarization is calculated:

\begin{equation}
\label{eq:Stot}
S_{tot}=(1-f_{np}) \int n(\alpha) \mathcal{S}(\alpha) d\alpha
\end{equation}

Including both these effects, we obtain excellent agreement with the experiment. This agreement with a time-averaged modulation-dressed state picture for the eigenlevels is notable as the period of the modulation is approximately only an order of magnitude shorter than the duration of the experiment \footnote{It should also be noted that modulating the Raman coupling heats the atoms. For $\Omega_M/\Omega_0=0.4$, this heating rate was measured to be less than $\approx2$ nK per ms. So given that our critical temperature is $\approx100$ nK and our experiments here last only 1.5 ms, this additional heating does not significantly harm our ability to see interference fringes or explore the spin-momentum locked ground modulation-dressed band.}.

The third verification of our interferometer came from tuning the contrast of the Stuekelberg interference by tuning $\Omega_C$, the gap size at the pair of avoided crossings. $\Omega_C$ partially determines $P_{LZ}$ and was tuned by varying $\Omega_M$. Thus changing $\Omega_M$ tunes the fraction of the BEC that splits into each leg of the interferometer with maximum contrast expected for $P_{LZ}=0.5$. Fig.~\ref{fig:varyOmegaM} (a) shows Stueckelberg interference for a few representative values of $\Omega_M$. Figure \ref{fig:varyOmegaM} (b) shows the spin contrast, defined as $M=(\mathcal{S}_{max}-\mathcal{S}_{min})/2$, for various values of $\Omega_M$. The results show how $\Omega_M$ can be used to control the Landau-Zener transitions and thus the interference fringe amplitude. 
The relationship between $\Omega_C$ and $\Omega_M$, calculated from the Floquet Hamiltonian exhibits a nearly linear dependence on $\Omega_M$. In fact, one finds that $\Omega_C\approx\Omega_M/2.27$ as shown in Fig.~\ref{fig:varyOmegaM}(c). Experimentally we find $\Omega_C=\Omega_M/2.3$ as the best estimate from our data, close to the theoretical calculation.

\section{Conclusions and Acknowledgements}
\label{sec_concl}
In summary, we explored the modulation-dressed bands of the SOC BEC created by modulating the Raman coupling strength. We observed the more complicated spin-momentum locking of the modulation-dressed band and engineered an atomic interferometer with the pair of avoided crossings between the modulation-dressed eigenlevels. Our measurements of Stueckelberg interference fringes agree with the theoretical analysis and thus confirms the treatment of the periodically modulated coupling. Interestingly, since the SOC is itself the result of dressing the single-particle dispersion with a Raman coupling, this can be considered as ``dressing'' the dressed states. This is another way to engineer novel light-induced synthetic gauge fields (for other examples, see Refs. \cite{Jimenez-Garcia2015,Zhang_NatSciRep_2013}). These initial experiments show the promise of this additional dressing, which offers new opportunities to study a novel SOC band-structures \cite{Lin_Nature_2011,Campbell2016}. For example, by choosing appropriate values of $\Omega_M$, $\Omega_0$, and $\delta$, we can realize $E_-(q)$ with three degenerate minima, in contrast to the 2 minima of $E_L(q)$. Such a novel SOC band and dispersion may uncover new physics of spinor and SOC BECs and deserves further exploration.

The research was supported in part by DURIP-ARO Grant No. W911NF-08-1-0265, the Miller Family Endowment, and a Purdue University OVPR Research Incentive Grant. A.J.O. also acknowledges support of the U.S. National Science Foundation Graduate Research Fellowship Program. C.Q. and C.Z. are supported by AFOSR (FA9550-16-1-0387) and
NSF (PHY-1505496). We thank Chris Greene, Li You, and Babak Seradjeh for helpful discussions.

\section{Appendix: Interferometry theory}
We use the matrix method to solve for the BEC eigenlevel population resulting from the BEC splitting, phase accumulation, and recombination. $\ket{\psi_{\pm}}$ indicates the wavefunction in the $E_{\pm}$ eigenlevels respectively, so the state of the BEC is expressed $\ket{\psi}=c_+ \ket{\psi_+}+c_- \ket{\psi_-}$. In operator notation, the state of the BEC is expressed
\begin{equation}
\ket{\psi} = \left( \begin{array}{c} c_+ \\ c_- \end{array} \right),
\end{equation}
and the beam splitters take the form
\begin{equation}
\hat{B}_{A} = \left( \begin{array}{ccc}
-\sqrt{1-P_{LZ}} & \sqrt{P_{LZ}} \\
\sqrt{P_{LZ}} & \sqrt{1-P_{LZ}} \end{array} \right)
\end{equation}
\begin{equation}
\hat{B}_{B} = \left( \begin{array}{ccc}
\sqrt{1-P_{LZ}} & \sqrt{P_{LZ}} \\
\sqrt{P_{LZ}} & -\sqrt{1-P_{LZ}} \end{array} \right)
\end{equation}
in which $P_{LZ}$ is the probability to make a diabatic transition in the modulation-dressed eigenlevels across the avoided crossing, and the negative signs on the diagonals account for phase shifts on the wavefunctions at each beam splitter \cite{Zeilinger_AJP_1981}. The phase difference accumulated by the components of the BEC can be accounted for by a phase operator defined by:
\begin{equation}
\hat{\Phi}(\phi) = \left( \begin{array}{ccc}
e^{i \phi/2} & 0 \\
0 & e^{-i \phi/2} \end{array} \right)
\end{equation}
where $\phi$ is the phase difference accumulated. Readout of the final state composition is done by Stern-Gerlach separation of the bare-$\ket{m_F}$ states when the BEC has crossed both $A$ and ${B}$ at a point when the $E_{\pm}$ eigenlevels match the bare states to better than $97 \%$, so that the spin polarization$=(N_{m_F=0}-N_{m_F=-1})/(N_{m_F=0}+N_{m_F=-1}) \approx (N_{\ket{+}}-N_{\ket{-}})/(N_{\ket{+}}+N_{\ket{-}})$. Thus, the readout of the spin polarization is given by 
\begin{equation}
\hat{S} = \left( \begin{array}{ccc}
1 & 0 \\
0 & -1 \end{array} \right).
\end{equation}
The final state after the beam splitter $A$, phase operator, and beam splitter $B$ is thus $\ket{\psi_f} = \hat{B}_B \hat{\Phi} \hat{B}_A \ket{\psi_i}$. The spin polarization reads $\bra{\psi_f} \hat{S} \ket{\psi_f}$, and when solved with $\ket{\psi_i}=\left( \begin{array}{c} 0 \\ 1 \end{array} \right)$ results in Eqn. (5) of the main text.

\bibliography{stueckPaper}

\end{document}